# Comparative Statistical Analysis of Prompt and Afterglow X-Ray Flares in Gamma-Ray Bursts: Insights into Extended Central Engine Activity

Yinuo Ma[1,2] and He Gao[1,2]
[1] Institute for Frontier in Astronomy and Astrophysics, Beijing Normal University, Beijing 102206, People's Republic of China; gaohe@bnu.edu.cn
[2] School of Physics and Astronomy, Beijing Normal University, Beijing 100875, People's Republic of China
*Received 2025 February 23; revised 2025 May 19; accepted 2025 June 2; published 2025 August 7*

## Abstract

Gamma-ray bursts (GRBs) are the most energetic phenomena in the Universe, characterized by prompt gamma-ray emission followed by multiwavelength afterglows. X-ray flares, observed during the afterglow phase, are generally believed to originate from the prolonged activity of the central engine, though direct evidence has been scarce. In this study, we present a comprehensive statistical analysis of X-ray flares from 315 GRBs observed by the Swift/X-ray Telescope over nearly two decades. We categorize flares into prompt flares (occurring during the prompt emission phase) and afterglow flares and compare their temporal and luminosity properties. Our analysis reveals that both types of flares exhibit similar morphological characteristics, with prompt flares being brighter and occurring earlier than afterglow flares. We find strong correlations between flare parameters, such as peak time, duration, and luminosity, which follow consistent patterns across both flare types. These findings suggest that X-ray flares, regardless of their timing, share a common origin in the central engine's activity. Our results imply that the central engine's activity duration extends beyond the prompt gamma-ray emission phase, highlighting the importance of considering X-ray flares when studying GRB progenitors and central engine properties. This work provides robust statistical evidence supporting the central engine origin of X-ray flares and underscores the need for future observations with missions like the Space-based multi-band astronomical Variable Objects Monitor and Einstein Probe to further elucidate the nature of GRB central engines.

*Unified Astronomy Thesaurus concepts:* Gamma-ray bursts (629)

*Materials only available in the* online version of record: machine-readable table

## 1. Introduction

Gamma-ray bursts (GRBs) are the most energetic phenomena in the Universe. Observationally, the significant characteristic of a GRB is the rapid increase and decrease in the gamma-ray band with short duration from several milliseconds to thousands of seconds (B. Zhang 2018). This bursty emission is defined as the prompt emission phase, and the prompt emission is produced by the internal dissipation processes of a relativistic jet (R. Sari & T. Piran 1997). It is generally believed that the duration of the prompt emission is related to the temporal behavior of the central engine (S. Kobayashi et al. 1997), such as the accretion timescale of black holes (BHs)/neutron stars or the release timescale of rotational and magnetic energy from magnetars (V. V. Usov 1992; S. E. Woosley 1993; B. Zhang & P. Mészáros 2001; W.-H. Lei et al. 2013; P. Kumar & B. Zhang 2015). After the prompt emission, there is a long-lasting multiwavelength afterglow, which is produced by the external dissipation processes as the relativistic jet interacts with the interstellar medium (B. Paczynski & J. E. Rhoads 1993; P. Mészáros & M. J. Rees 1997; A. Panaitescu & P. Kumar 2002; H. Gao et al. 2013). The timescale of the afterglow is observed for a few minutes, several weeks, months, or even years and is primarily determined by factors such as the velocity of the relativistic jet and the density of the interstellar medium.

The rapid slew capability of the narrow-field instrument X-ray Telescope (XRT) on board the Swift spacecraft (N. Gehrels et al. 2004) allowed afterglow to be observed at very early times, which significantly accumulated the early afterglow samples with measured redshift. Especially, a canonical X-ray afterglow light curve characterized by five components is revealed by B. Zhang et al. (2006). The interesting phenomena clearly exceeding the background power-law decay component are called X-ray flares, and they have steep rising and decaying indices. In the Swift satellite observations, one third of the X-ray afterglows have X-ray flares (G. Chincarini et al. 2010). The peak times of X-ray flares typically occur between 100 and $10^6$ s after the GRB trigger, with the majority concentrated in the range of 100–1000 s (S.-X. Yi et al. 2016). Such large slopes of X-ray flares cannot be explained by external shocks, and a more natural explanation for them is dominated by internal dissipation related to a long-lived central engine (D. N. Burrows et al. 2005; P. Romano et al. 2006). The main theories include the fragmentation and subsequent accretion from a rapidly rotating collapsar (A. King et al. 2005), differential rotating millisecond pulsars after the binary mergers (Z. G. Dai et al. 2006), fragmentation of a hyperaccreting accretion disk (R. Perna et al. 2006), transition of the accretion modes (D. Lazzati et al. 2008), He-synthesis-driven disk winds (W. H. Lee et al. 2009), precessing jets (T. Liu et al. 2010; S.-J. Hou et al. 2014), propagation instabilities in jets (D. Lazzati et al. 2011), magnetic coupling in a BH hyperaccretion disk (Y. Luo et al. 2013), jets driven by the Blandford–Znajek mechanism (W.-H. Lei et al. 2013; W. Xie et al. 2017; S.-X. Yi et al. 2021), and the fallback accretion on a magnetar propeller (S. L. Gibson et al. 2017, 2018; W.-Y. Yu et al. 2024). The occurrence of X-ray flares indicates an extension in the timescale of the central engine. Consequently, it is widely argued that the duration of the prompt gamma-ray emission may not reflect the timescale of the intrinsic central engine activity. Some authors have proposed redefining the burst duration by simultaneously considering the characteristics of gamma-ray and X-ray light curves (B.-B. Zhang et al. 2014; M. Boër et al. 2015; H. Gao et al. 2017). The Einstein Probe (EP) observations of EP240315a also clearly show that the prompt X-ray duration is longer than that of the gamma rays, when the Wide-field X-ray Telescope captured the earlier engine activation and extended late engine activity (Y. Liu et al. 2025). Therefore, the study of X-ray flares is of significant importance to determine the temporal behavior of the central engine and to understand the properties of its progenitor.

How to confirm that X-ray flares are produced by the central engine is an essential question. Previous studies have focused on comparing the characteristics of X-ray flares and the prompt emission to reveal the relation between X-ray flares and the central engine (A. D. Falcone et al. 2007; R. Margutti et al. 2010). Statistical studies of X-ray flares have found that some GRBs have only one flare, while some have multiple flares (G. Chincarini et al. 2007). The flare spectra require a more complex spectral model than the simple power law of the afterglow radiation (A. D. Falcone et al. 2006). In particular, flares follow a hard-to-soft evolution, which is the same as the prompt emission (L. A. Ford et al. 1995; L.-Z. Lü et al. 2022). G. Chincarini et al. (2007) found the same distribution of intensity ratios between successive prompt emission pulses and X-ray flares. X. Z. Chang et al. (2021) revealed that multiflare GRBs and multipulse GRBs have similar positive lag–duration correlations. The evidence mentioned above indicates that X-ray flares have the same physical origin as prompt emission, which are directly powered by the GRB central engine. However, some works have revealed that there is no evident correlation between the different parameters of X-ray flares and the prompt emission (G. Chincarini et al. 2007, 2010; J. Saji et al. 2023). Therefore, a more direct method that reflects the origin of the flares would be essential.

Previous works have already found that a few X-ray flares can be simultaneously detected by the Burst Alert Telescope and the XRT on board the Swift spacecraft (e.g., F.-K. Peng et al. 2014; W.-Y. Yu et al. 2024). Based on the results of joint temporal and spectral analyses with gamma rays, it is conclusively determined that this type of flare originates from the activity of the central engine (N. R. Butler & D. Kocevski 2007; R. Margutti et al. 2010). Henceforth, we refer to these flares as "prompt flares." As the operational time of Swift increases, the samples of X-ray flare accumulate year by year. We analyzed data from Swift/XRT over the past two decades and found that >20% of the flares were prompt flares. In this work, we conducted a detailed analysis to compare and contrast various properties of flares occurring during the afterglow phase with those of prompt flares. This involves assessing whether flare parameters, particularly the distributions of their rise and decay slopes, are consistent and exploring the coherence of relationships between parameters like flare peak times and durations, as well as peak times and luminosities. Our objective, from a statistical data analysis standpoint, is to definitively ascertain whether flares during the afterglow phase, akin to prompt flares, indeed originate from the activity of the central engine.

## 2. Sample Selection and Methodology

We search the 0.3–10 keV X-ray light curves of GRBs from the website of Swift/XRT (P. A. Evans et al. 2007, 2009) between 2005 January and 2024 August. In the past two decades, there have been nearly 2000 GRBs, and we have filtered these samples based on the following three criteria:

(1) The flare exhibits a relatively complete structure, including three distinct phases: rise, peak, and decay phase.

(2) A feature is classified as a flare only when its flux exceeds twice the afterglow background level ($F_{\text{flare}} > 2F_{\text{background}}$), whereas smaller fluctuations below this threshold are not identified as flares.

(3) The underlying afterglow emission has enough observational data so that it can be fitted with a power-law function.

After visually examining, we obtained 315 GRBs with 701 flares, among which 242 flares of 109 GRBs have redshift measurements. For these GRBs, we systematically divided their afterglow light curves into two distinct components: (1) the underlying smooth afterglow background, modeled as a power-law decay, and (2) the X-ray flare components superimposed on the continuum. The separation was performed through the following fitting process: First, we modeled the underlying afterglow background by fitting either a single power law or a broken power law to the quiescent (nonflaring) portions of the light curve using nonlinear least-squares minimization. After obtaining the best-fit background model, we subtracted it from the observed light curve to isolate the flare components. Then, we fitted each individual pulse within the flaring episode with a smoothly broken power law:

$$F(t) = A \left(\frac{t}{t_{\text{b}}}\right)^{-\alpha_1} \left\{\frac{1}{2}\left[1 + \left(\frac{t}{t_{\text{b}}}\right)^{1/\Delta}\right]\right\}^{(\alpha_1 - \alpha_2)\Delta}, \quad (1)$$

where $t_{\text{b}}$ is the break point, $A$ is the model amplitude at the break point $t_{\text{b}}$, $\alpha_1$ is the power-law index for $t \ll t_{\text{b}}$, $\alpha_2$ is the power-law index for $t \gg t_{\text{b}}$, and $\Delta$ represents the smoothness of the peak. This method is similar to the fitting of G. Chincarini et al. (2007, 2010) and S.-X. Yi et al. (2016).

The fitting result of GRB 140506A, as shown in Figure 1, serves as an example. We fit the afterglow light curve of GRB 140506A with three smoothly broken power laws (blue dotted lines) and a power law (green dashed line). The time parameters of flares are defined as follows. The start time $T_{\text{start}}$ and the end time $T_{\text{end}}$ of a flare as the two intersection points between the flare fitting curve and the afterglow continuum light curve. The duration of the flare $T_{\text{duration}}$ equals the end time $T_{\text{end}}$ minus the start time $T_{\text{start}}$. The rise time $T_{\text{rise}}$ is defined as the time from the start to the peak, and the decay time $T_{\text{decay}}$ is defined as the time from the peak to the end. The flux corresponding to the peak time $T_{\text{peak}}$ is denoted as $F_{\text{peak}}$, and the fluence $S_{\text{F}}$ is the integral of the smoothly broken power law over the time interval from the start to the end of the flare. All the flare parameters should be transferred from the observer frame to the source frame if they have measured redshifts. Time $T$ of the observer frame is converted to $T/(1 + z)$ in the source frame, where $z$ is the redshift of a GRB. Consequently, the isotropic energy of the X-ray flare can be derived as $E_{\text{x,iso}} = 4\pi D_{\text{L}}^2 S_{\text{F}}/(1 + z)$. $D_{\text{L}}$ is the luminosity distance and is calculated by

$$D_{\text{L}}(z) = \frac{c(1 + z)}{H_0} \int_0^z \frac{dz'}{\sqrt{\Omega_{\text{M}}(1 + z')^3 + \Omega_\Lambda}}, \quad (2)$$

where $H_0$ is the Hubble constant, $\Omega_{\text{M}}$ is the dimensionless matter density, and $\Omega_\Lambda$ is the dark energy density parameters of the Universe. In this paper, $H_0 = 71 \text{ km s}^{-1} \text{ Mpc}^{-1}$, $\Omega_M = 0.30$, and $\Omega_\Lambda = 0.70$. Additionally, the peak isotropic luminosity and mean isotropic luminosity can be derived from $L_{p,iso} = 4\pi D_L^2 F_p$ and $L_{x,iso} = (1 + z)E_{x,iso}/T_{duration}$. The k-correction is not considered in this paper.

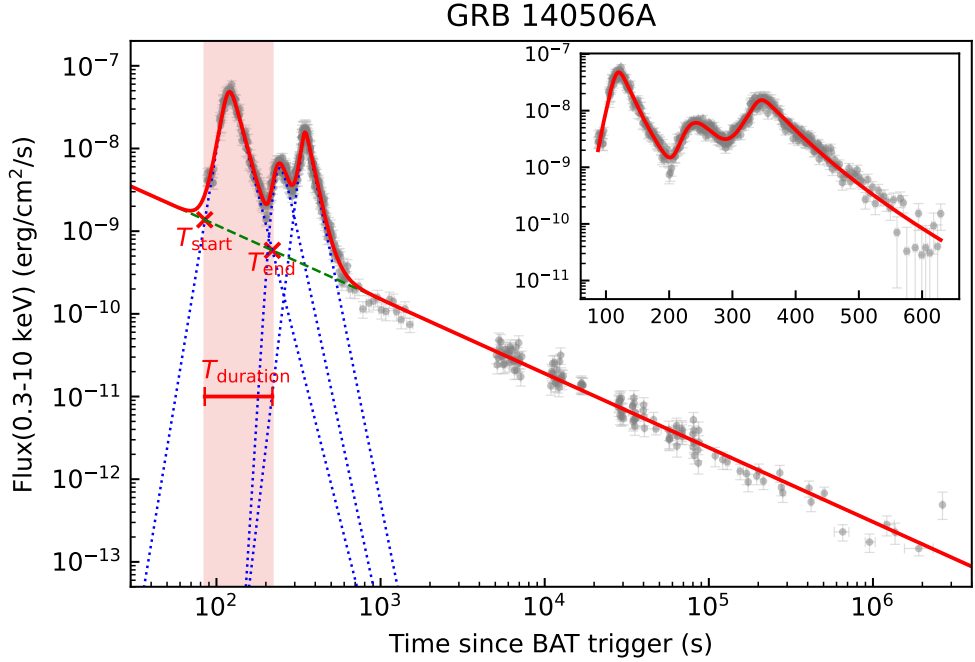

**Figure 1.** The fitting plot of GRB 140506A serves as an example. The green dashed line represents the fit to the underlying afterglow background, the blue dotted lines show the best fit to the flares, and the red solid line represents the superposition of them. The markers "x" denote the start and end times of the first flare, and the red shaded area indicates the duration of the first flare. The detail of flare fittings is drawn in the embedded plot.

To determine the variations of parameters of flares, such as $T_{start}$ and $T_{end}$, a Gaussian distribution was assumed to generate 5000 simulated X-ray light curves using the Monte Carlo method like J. Saji et al. (2023). Let the flux at times $(t_1, …, t_n)$ be $(flux_1, …, flux_n)$, with their respective errors being $(\sigma_{t_1},…,\sigma_{t_n})$ and $(\sigma_{flux_1},…,\sigma_{flux_n})$. We generate $N = 5000$ simulated samples $(t_1^{(i)},…,t_n^{(i)})$ and $(flux_1^{(i)},…,flux_n^{(i)})$ for $i = 1, …, N$, where $t_j^{(i)} \sim \mathcal{N}(t_j, \sigma_{t_j}^2)$, $flux_j^{(i)} \sim \mathcal{N}(flux_j, \sigma_{flux_j}^2)$ for $j = 1, …, n$. That is, the original data points of the light curve serve as the mean of the Gaussian distribution, with their errors representing the standard deviation. Then, a smoothly broken power law is used to fit the light curve of the flare in the same manner. The mean and standard deviation of the distribution of the start times $T_{start}^{(i)}$ and end times $T_{end}^{(i)}$, $i = 1,…,N$, as derived from the fitting, are taken as $T_{start}$, $T_{end}$, and their statistical errors. Other parameters are also acquired through a similar approach. The fitting parameters of the flares are shown in Table 1.

The duration of the prompt emission of a GRB is commonly defined as $T_{90}$, which refers to the time interval during which 5%–95% of the total gamma-ray fluence is detected by the instrument (C. Kouveliotou et al. 1993; P. Kumar & B. Zhang 2015; B. Zhang 2018). Here, we define an X-ray flare with $T_{peak} < T_{90}$ end time as a "prompt flare." We have identified 183 prompt flares, among which 69 flares have redshifts. The remaining 518 flares (with 173 having redshifts) are subsequently referred to as "afterglow flares." We plotted GRBs with prompt flares on the Amati relation diagram (the correlation between the GRB isotropic gamma-ray prompt emission energy $E_{\gamma,iso}$ and the rest-frame peak energy $E_{peak,z}$, as shown in Figure 2) and found that these sources are all typically long GRBs without any particular peculiarity. The reason why these sources exhibit prompt flares should be purely due to the early availability and commencement of observations by the XRT. Therefore, if the properties of afterglow flares and prompt flares are consistent, it would universally indicate that they should share the same origin. With a substantial sample of flares with known redshifts, we are able to perform robust statistical analysis. The following comparisons are based exclusively on these redshift-confirmed flares to ensure physically meaningful evaluation of their intrinsic properties.

### 3. Data Analysis

#### 3.1. Distributions of X-Ray Flare Parameters

Figure 3 shows the histogram distributions of the temporal parameters for these two types of flares with measured redshifts, and the flare parameters are transferred from the observer frame to the source frame. Figures 3(a), (b), and (c) are histograms of the start time $T_{start}$, end time $T_{end}$, and peak time $T_{peak}$ of the flares, respectively. The smoothed versions of the distributions obtained via kernel density estimation (KDE) using Gaussian kernels reveal that the peaks for $T_{start}$, $T_{end}$, and $T_{peak}$ of prompt flares all occur earlier than those of afterglow flares. Figure 3(d) presents the distributions of the flare durations $T_{duration}$. The durations of afterglow flares are distributed across a wide range, from 1 to $10^5$ s, with a concentration between 10 and 1000 s. The distribution of durations for prompt flares is not significantly different from that of afterglow flares, although prompt flares lack those with very long durations. This is because the duration of the prompt emission of a GRB is inherently short compared with the afterglow, and naturally, the prompt flares will not last very long either. To quantitatively compare whether these two categories of flares belong to the same distribution, we conduct Kolmogorov–Smirnov (K-S) tests on these parameters for the flares with measured redshifts.

Table 1
Fitting Results of X-Ray Flares

| GRB | z | $T_{start}$ (s) | $T_{end}$ (s) | $T_{peak}$ (s) | $F_{peak}$ ($10^{-10}$ erg cm$^{-2}$ s$^{-1}$) | $S_F$ ($10^{-8}$ erg cm$^{-2}$) | $\alpha_1$ | $\alpha_2$ | $L_{p,iso}$ ($10^{48}$ erg s$^{-1}$) | $E_{x,iso}$ ($10^{50}$ erg) |
|---|---|---|---|---|---|---|---|---|---|---|
| 050406 | ⋯ | 111.964 ± 35.839 | 514.940 ± 327.573 | 211.776 ± 12.454 | 1.328 ± 0.236 | 2.337 ± 7.812 | −6.584 ± 8.254 | 7.817 ± 10.680 | ⋯ | ⋯ |
| 050502B | ⋯ | 400.627 ± 41.583 | 1420.302 ± 114.166 | 706.993 ± 10.047 | 32.774 ± 1.023 | 78.467 ± 2.638 | −11.723 ± 2.148 | 10.846 ± 1.081 | ⋯ | ⋯ |
| 050502B | ⋯ | 27070.423 ± 3172.011 | 37497.396 ± 9592.148 | 30449.477 ± 2799.286 | 0.011 ± 0.008 | 0.605 ± 0.427 | −10.257 ± 10.828 | 7.323 ± 6.723 | ⋯ | ⋯ |
| 050502B | ⋯ | 54554.422 ± 5470.854 | 108185.535 ± 9130.353 | 78485.115 ± 1968.420 | 0.018 ± 0.009 | 4.346 ± 1.237 | −4.895 ± 1.947 | 7.320 ± 1.990 | ⋯ | ⋯ |
| 050607 | ⋯ | 243.728 ± 24.854 | 682.447 ± 105.400 | 303.251 ± 8.450 | 8.033 ± 1.303 | 7.254 ± 0.876 | −27.944 ± 15.278 | 7.391 ± 3.316 | ⋯ | ⋯ |
| 050712 | ⋯ | 144.488 ± 17.006 | 300.870 ± 42.943 | 218.327 ± 1.835 | 4.545 ± 1.166 | 2.379 ± 0.546 | −6.669 ± 1.974 | 11.305 ± 4.830 | ⋯ | ⋯ |
| 050712 | ⋯ | 203.251 ± 17.340 | 596.185 ± 153.178 | 258.344 ± 4.792 | 5.452 ± 2.321 | 4.877 ± 1.007 | −13.722 ± 3.874 | 5.285 ± 2.385 | ⋯ | ⋯ |
| 050712 | ⋯ | 380.997 ± 45.974 | 625.868 ± 77.788 | 494.670 ± 7.931 | 2.584 ± 1.120 | 2.216 ± 1.305 | −12.754 ± 8.759 | 15.657 ± 7.300 | ⋯ | ⋯ |
| 050712 | ⋯ | 871.110 ± 56.956 | 1383.861 ± 536.020 | 992.799 ± 26.243 | 0.447 ± 0.221 | 0.876 ± 0.408 | −12.495 ± 5.294 | 7.295 ± 3.634 | ⋯ | ⋯ |
| 050713A | ⋯ | 91.567 ± 9.118 | 158.203 ± 22.578 | 109.974 ± 1.973 | 77.856 ± 16.121 | 17.860 ± 4.364 | −29.973 ± 22.712 | 12.741 ± 5.424 | ⋯ | ⋯ |
| 050713A | ⋯ | 155.184 ± 3.727 | 222.274 ± 7.474 | 166.945 ± 5.288 | 8.572 ± 3.016 | 2.625 ± 1.202 | −38.141 ± 19.763 | 15.499 ± 14.194 | ⋯ | ⋯ |

(This table is available in its entirety in machine-readable form in the online article.)

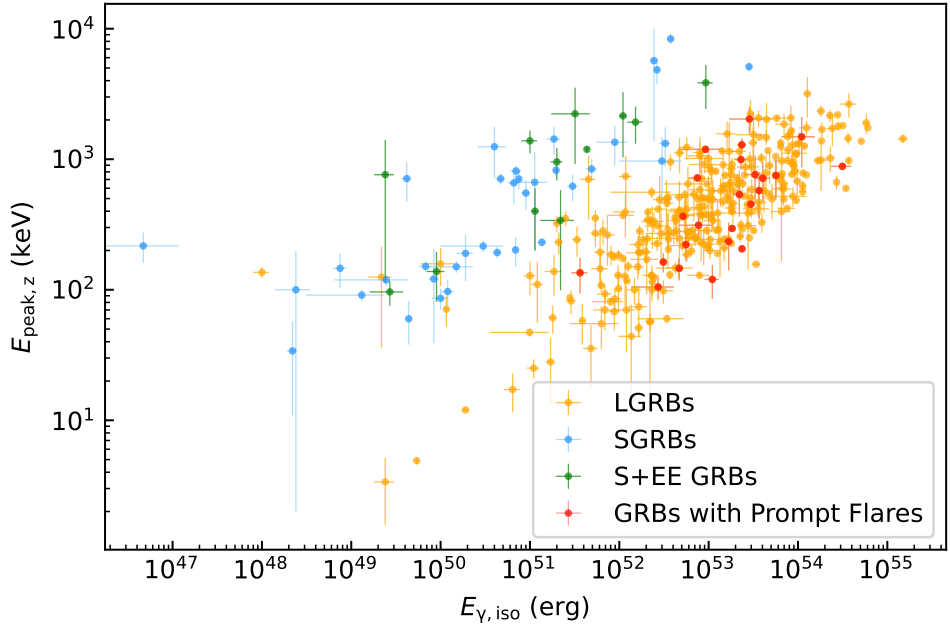

**Figure 2.** $E_{peak,z}$–$E_{\gamma,iso}$ (Amati) relation with known redshift data (L. Lan et al. 2023) is shown. The orange points, blue points, green points, and red points represent the long GRBs, short GRBs, short GRBs with an extended emission, and GRBs with prompt fares in this paper, respectively.

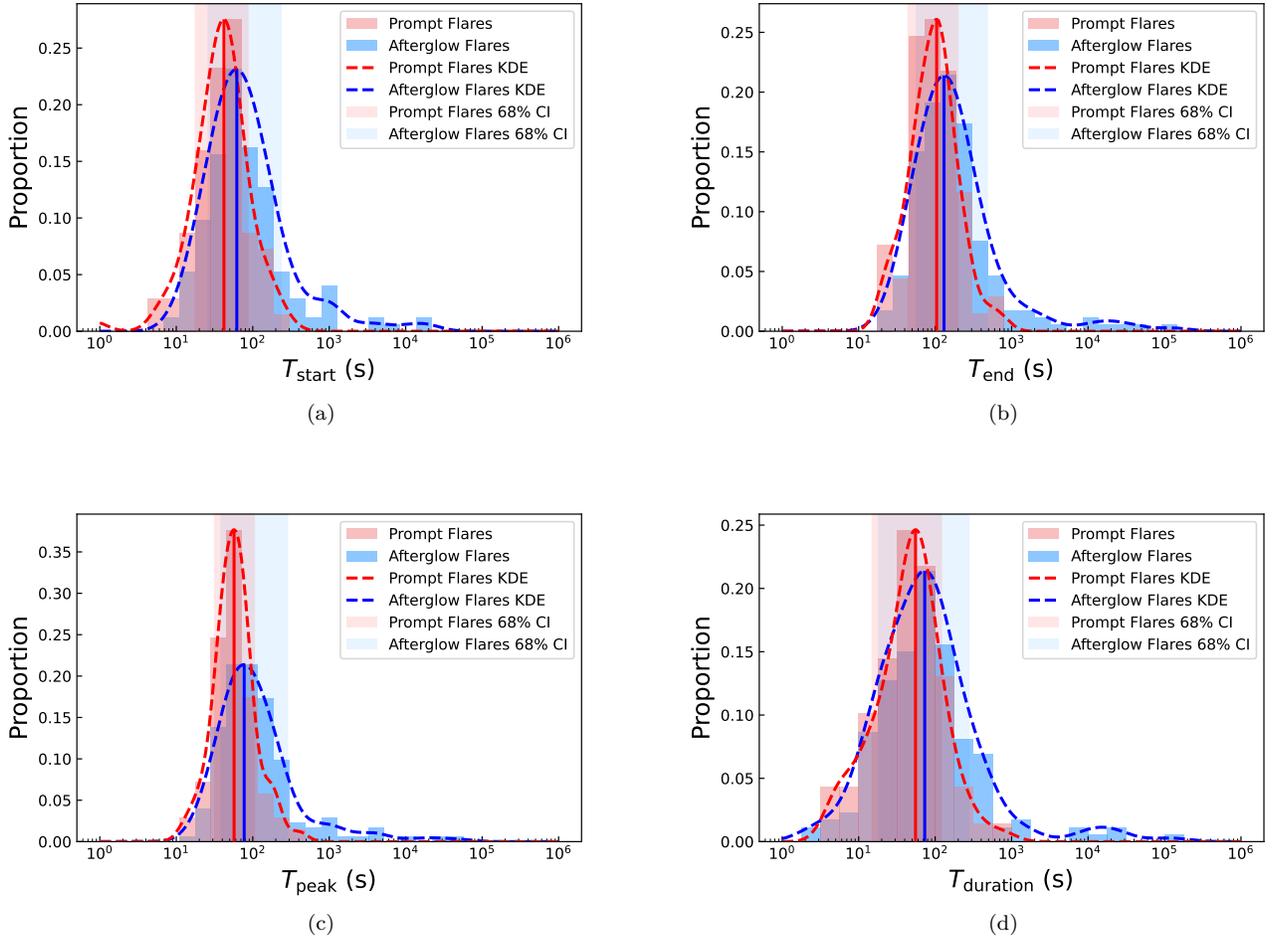

**Figure 3.** Histogram distributions of temporal parameters for the two X-ray flare types with known redshifts: (a) start time $T_{start}$, (b) end time $T_{end}$, (c) peak time $T_{peak}$, and (d) duration $T_{duration}$ are shown. The dark-red shaded region corresponds with the distribution of prompt flares, while the dark-blue shaded region represents afterglow flares. The red and blue dashed lines represent the KDE for prompt flares and afterglow flares, respectively, while the solid lines indicate the peaks of the KDE distributions. The light-colored shaded areas show their 68% confidence intervals. The same color scheme is used hereafter.

Table 2
K-S Tests of X-Ray Flares

| Parameters | $T_{start}$ | $T_{peak}$ | $T_{end}$ | $T_{duration}$ | $T_{ratio}$ | $\Delta T/T$ |
|---|---|---|---|---|---|---|
| P-values | $6.974 \times 10^{-6}$ | $2.055 \times 10^{-5}$ | $2.681 \times 10^{-4}$ | 0.007 | 0.087 | 0.657 |
| Parameters | $\alpha_1$ | $\alpha_2$ | $F_p$ | $S_F$ | $L_{p,iso}$ | $E_{x,iso}$ |
| P-values | 0.329 | 0.138 | $5.124 \times 10^{-11}$ | $1.494 \times 10^{-4}$ | $2.478 \times 10^{-5}$ | 0.003 |

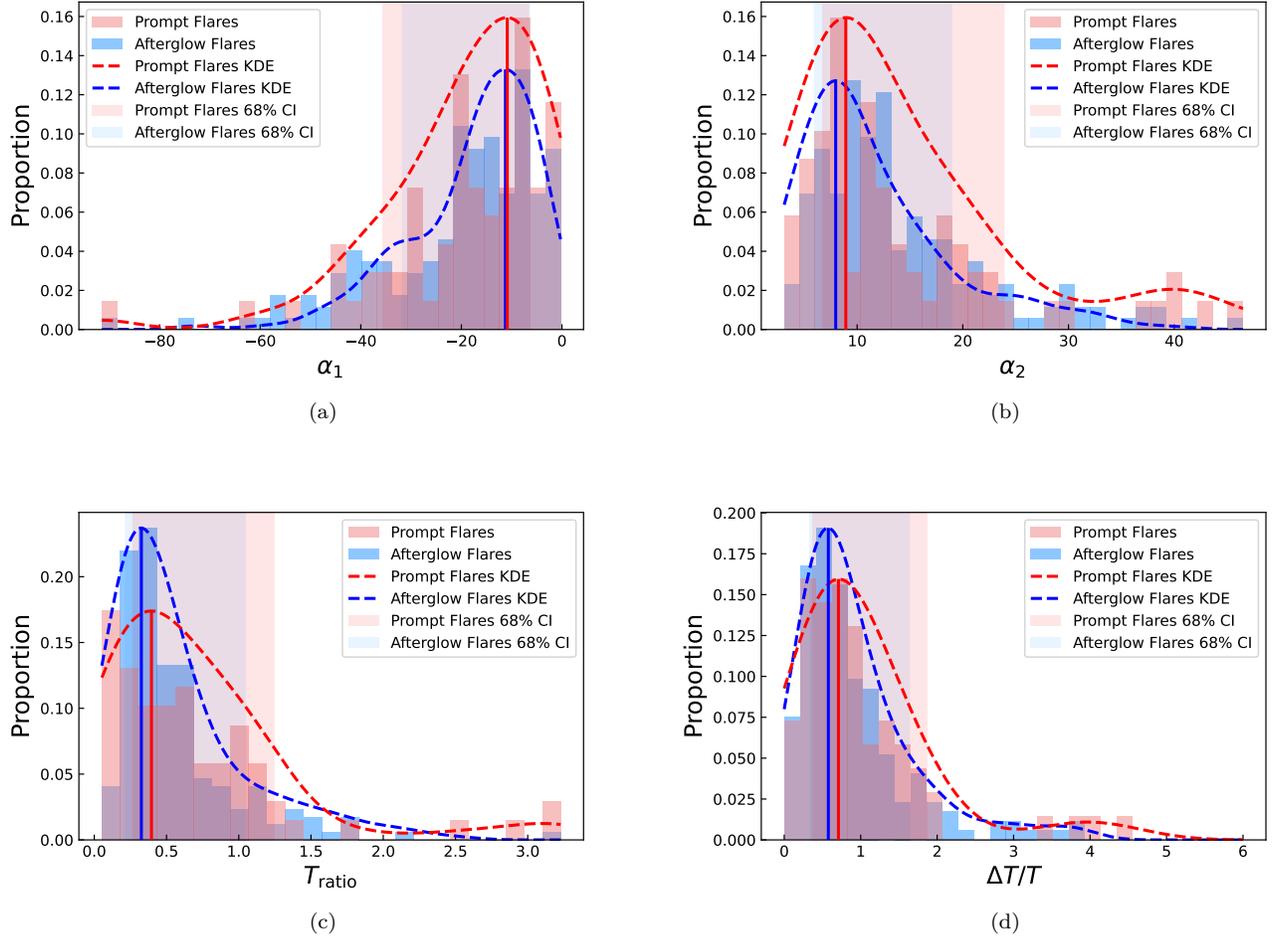

**Figure 4.** Histogram distributions of X-ray flare morphological parameters: (a) flare rise slope $\alpha_1$, (b) flare decay slope $\alpha_2$, (c) ratio of flare rise time to decay time $T_{ratio}$, and (d) ratio of flare duration to peak time $\Delta T/T$ are shown. Color conventions follow those in Figure 3.

The results of the K-S tests are shown in Table 2. The P-values of the K-S tests for the temporal parameters are less than 0.05, leading to the rejection of the assumption that the occurrence times of prompt and afterglow flares originate from the same distribution, as discussed above.

Figure 4 shows the histogram distributions of the morphological parameters of X-ray flares. Figures 4(a) and (b) show the distributions of the flares rise slope $\alpha_1$ and decay slope $\alpha_2$, respectively, which indicate the flare steepness. Figure 4(c) draws the distribution of the ratio of the flare rise time to the decay time $T_{ratio}$, which reflects the symmetry of the flares. Figure 4(d) illustrates the distribution of the ratio of the flare duration to the peak time $\Delta T/T$, which indicates the sharpness or flatness of the flares. To compare the consistency of their morphological feature distributions, we also employ the K-S tests. The P-values of the K-S tests for $\alpha_1$, $\alpha_2$, $T_{ratio}$, and $\Delta T/T$ are 0.329, 0.138, 0.087, and 0.657, respectively. All of these P-values are greater than 0.05, indicating that the two types of flares are very likely to have the same morphology.

Figures 5(a) and (b) are histograms of the peak flux $F_{peak}$ and fluence $S_F$ of the flares, respectively. The peak values of KDE for $F_{peak}$ and $S_F$ of prompt flares are higher than those of afterglow flares. Similarly, it can be observed that the peak luminosity $L_{p,iso}$ (as shown in Figure 5(c)), average luminosity $L_{x,iso}$ (as shown in Figure 5(d)), and isotropic energy $E_{x,iso}$ (as shown in Figure 5(e)) of prompt flares are brighter than those of afterglow flares. This may be because prompt flares occur earlier than afterglow flares. The observations are consistent with the findings presented in S.-X. Yi et al. (2016), which suggest that flares occurring later are generally dimmer. The P-values of the K-S tests for these parameters are also less than 0.05, quantitatively demonstrating that their luminosity distributions are different.

Overall, by comparing the distributions of the parameters for the two types of flares, we find that the morphological characteristics of these two categories of flares are quite similar. Additionally, the prompt flares are brighter than the afterglow flares, which is consistent with the regularity that flares occurring earlier tend to be brighter.

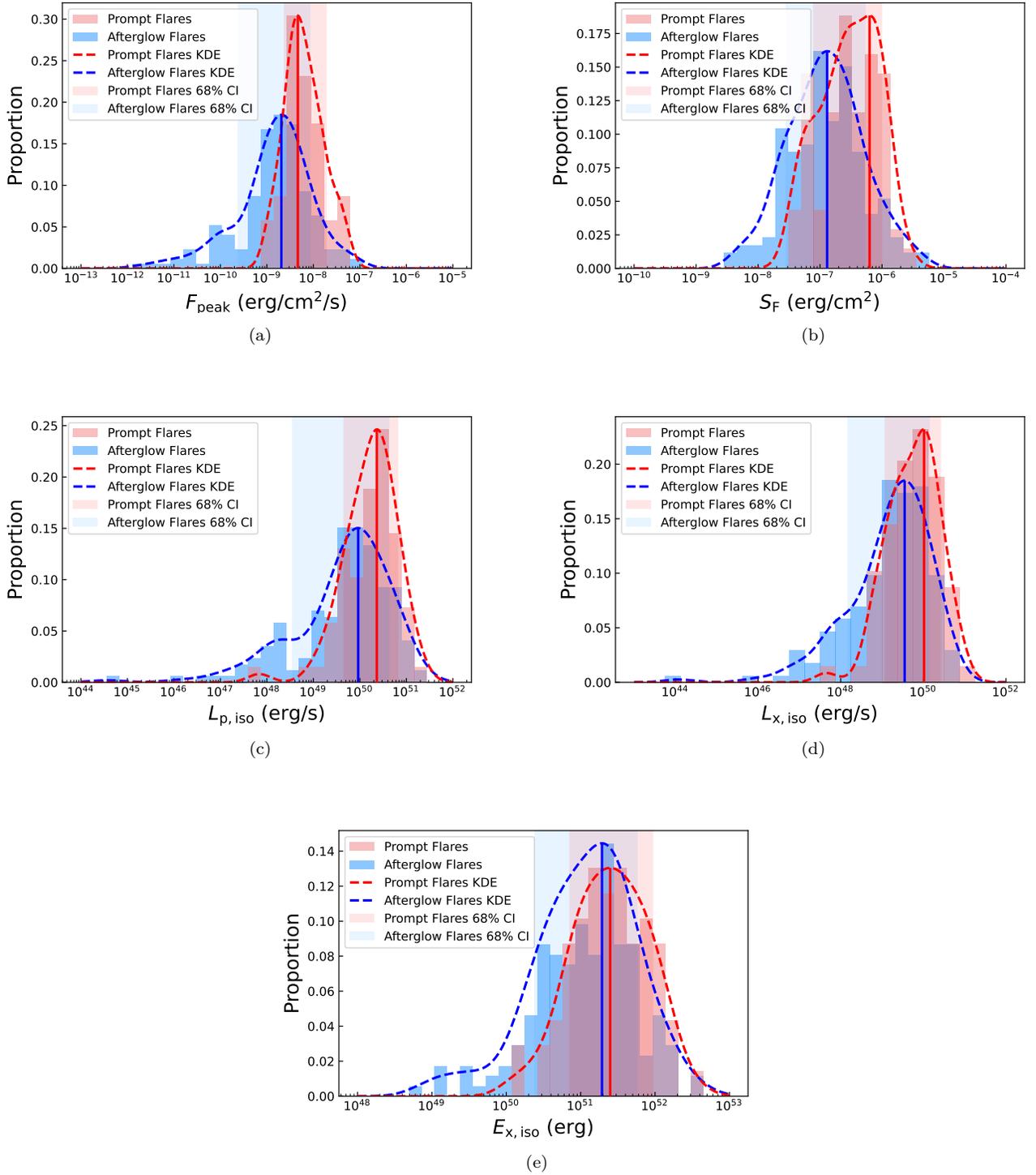

**Figure 5.** Histogram distributions of X-ray flare: (a) peak flux $F_{\rm peak}$, (b) fluence $S_{\rm F}$, (c) peak luminosity $L_{\rm p,iso}$, (d) average luminosity $L_{\rm x,iso}$, and (e) isotropic energy $E_{\rm x,iso}$ are shown. Color conventions follow those in Figure 3.

For prompt flares and afterglow flares that occur at the same time, will their luminosity and energy distributions differ? To answer this question, we selected flares with $T_{\rm peak} < 200$ s to ensure that the temporal distributions of these two types of flares are identical. As shown in Figure 6, their $L_{\rm p,iso}$ and $E_{\rm x,iso}$ distributions appear similar in this case. Quantitatively, the P-values from the K-S test for both are greater than 0.05 (Table 3). The P-values for $F_{\rm peak}$ and $S_{\rm F}$ being less than 0.05 are due to observational effects. Only the consistency in luminosity and energy can reflect the fundamentally identical origin of these two types of flares.

### 3.2. Correlation Analysis

To clearly demonstrate the correlations between flare parameters, the Spearman correlation matrix of the various flare parameters with redshift is shown in Figure 7. In the figure, there are positive correlations among the temporal parameters and negative correlations between luminosities and times, but no significant correlations between isotropic energy and times. It is also evident that there are clear positive correlations between isotropic energy and luminosities. The correlations mentioned above are consistent with previous studies (G. Chincarini et al. 2007, 2010; S.-X. Yi et al. 2016;

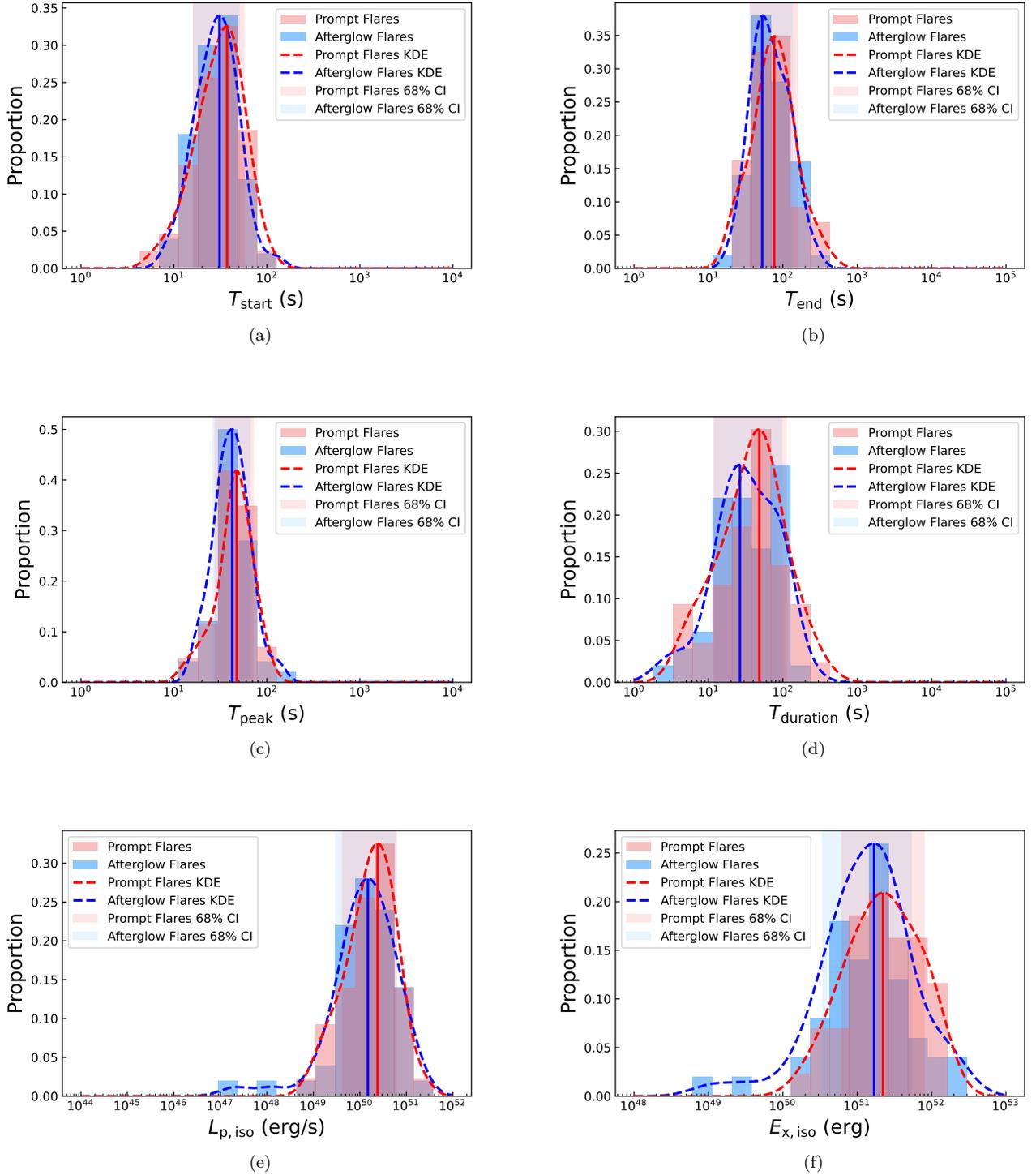

**Figure 6.** Histogram distributions of X-ray flare with $T_{\rm peak} < 200$ s: (a) start time $T_{\rm start}$, (b) end time $T_{\rm end}$, (c) peak time $T_{\rm peak}$, (d) duration $T_{\rm duration}$, (e) peak luminosity $L_{\rm p,iso}$, and (f) isotropic energy $E_{\rm x,iso}$ are shown. Color conventions follow those in Figure 3.

**Table 3**
K-S Tests of X-Ray Flares with $T_{\rm peak} < 200$ s

| Parameters | $T_{\rm start}$ | $T_{\rm peak}$ | $T_{\rm end}$ | $T_{\rm duration}$ | $T_{\rm ratio}$ | $\Delta T/T$ |
|---|---|---|---|---|---|---|
| P-values | 0.619 | 0.449 | 0.520 | 0.597 | 0.822 | 0.421 |
| Parameters | $\alpha_1$ | $\alpha_2$ | $F_{\rm p}$ | $S_{\rm F}$ | $L_{\rm p,iso}$ | $E_{\rm x,iso}$ |
| P-values | 0.349 | 0.224 | 0.012 | 0.026 | 0.655 | 0.169 |

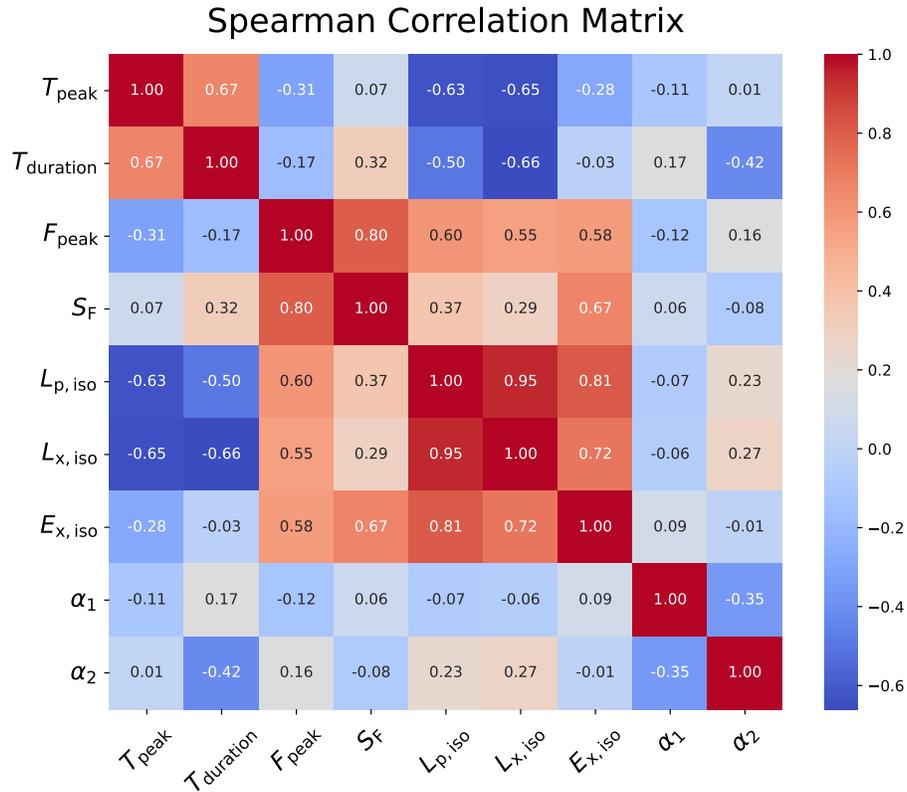

**Figure 7.** The Spearman correlation matrix of the various flare parameters with redshift is shown. The color gradient ranges from blue (strong negative correlation) to red (strong positive correlation).

Y.-R. Shi et al. 2022). Moreover, the correlations between parameters can be directly displayed by plotting scatter figures. Therefore, we select parameter pairs with Spearman correlation coefficients greater than 0.6 and perform logarithmic linear regression analyzes using linear least-squares regression fitting, conducting separate analyses for prompt flares and afterglow flares to determine whether they share the same underlying regularity. The results of the logarithmic linear regression analysis are shown in Table 4.

The left panel of Figure 8 shows a clear positive correlation between the duration $T_{\rm duration}$ and the peak time $T_{\rm peak}$ of the two types of flares with measured redshift, indicating that the later a flare occurs, the longer its duration tends to be. The slopes and intercepts of the log-linear correlations for the two types of flares, within their $1\sigma$ error ranges (determined by Markov Chain Monte Carlo, MCMC, simulation), exhibit significant overlap. This suggests that they should follow the same regularity, as shown in the right panel of Figure 8.

The top panel and the middle panel of Figure 9, respectively, illustrate that there are negative correlations between the peak and average luminosities of flares and their peak times, meaning that the flares occurring later are generally dimmer, which is consistent with the results from previous surveys. Moreover, the fitting parameters for these two types of flares overlap within the $1\sigma$ range, suggesting that they may have the same physical origin. There is also a negative correlation between the average luminosity and the duration, as shown in the bottom panel of Figure 9, indicating that flares with longer durations tend to be dimmer. The intercepts obtained from the fitting for these two types of flares overlap within the $1\sigma$ range, while although the slopes do not overlap within the $1\sigma$ range, they do overlap within the $2\sigma$ range.

The tight positive correlation between average luminosity and peak luminosity is shown in the top panel of Figure 10, while positive correlations between isotropic energy and fluence, average luminosity, and peak luminosity are exhibited in Figures 10 and 11, respectively. The three correlations in Figure 10 overlap within the $1\sigma$ range. However, the slopes and intercepts barely overlap in the top panel of Figure 11, which is due to the fact that the durations of the late flares cannot be well determined, and late relatively faint flares are drowned in the afterglow background. Therefore, the late flares we have selected are all brighter ones with long durations, which leads to some flares with larger energy that are significantly above the fitting line in the top panel of Figure 11. If we only consider early flares, i.e., flares with peak times less than 1000 s, this problem can be avoided, as shown in the bottom panel of Figure 11. A similar issue also exists in the positive correlation between fluence and peak flux, as shown in the top panel of Figure 12. The late flares have low peak flux, but their durations are very long, resulting in flares located in the upper-left corner of the fitting line. Similarly, when we consider only the early flares, the correlations between these two types of flares can also overlap. We also perform K-S tests on these early flares. Except for the $P$-value of duration being greater than 0.05, the other results are the same as those of full flare sample.

In summary, the consistencies in the correlations between the parameters of these two types of X-ray flares indicate that they follow the same regularity, further suggesting that both originate from the activity of the central engine.

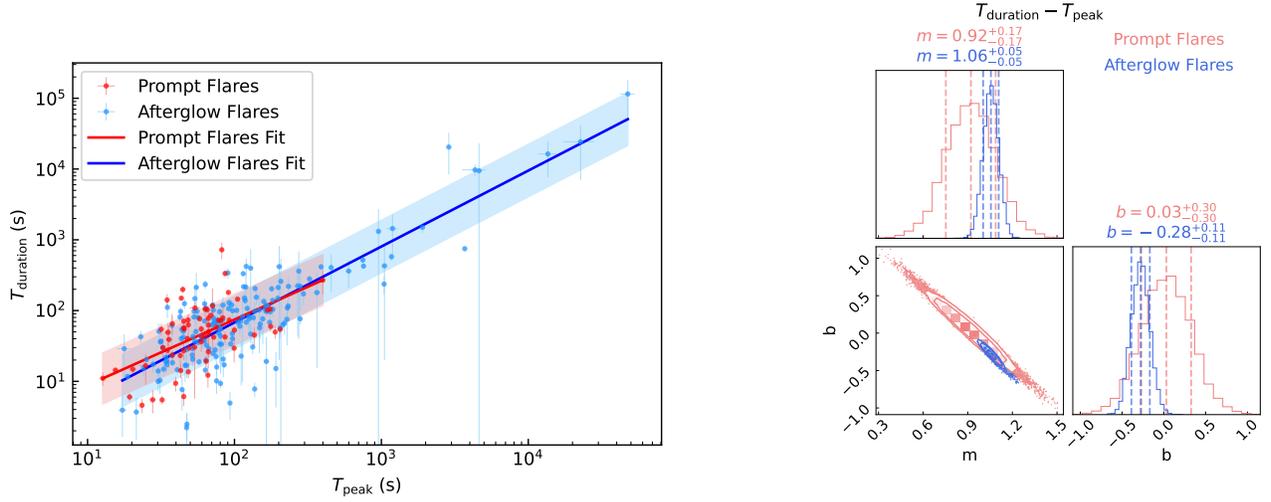

**Figure 8.** The correlation between the duration $T_{duration}$ and peak time $T_{peak}$ of GRB X-ray flares with measured redshift is shown. Left panel: the red points and blue points represent prompt and afterglow flares, respectively. The red solid line and the blue solid line represent the best log-linear regression fits for prompt and afterglow flares, respectively. The shaded red and blue regions correspond the $1\sigma$ deviations of the data from the fit. The same color scheme is used hereafter. Right panel: the corner plot of $T_{duration}$ and $T_{peak}$ with measured redshift by MCMC simulation. Red and blue represent prompt and afterglow fares, respectively. The same color scheme is used hereafter.

Table 4
Results of Logarithmic Linear Regression Analysis

| Correlations | Types | Expressions | R | P |
|---|---|---|---|---|
| $T_{duration}$ ($T_{peak}$) | Prompt Flares | $\log T_{duration} = (0.03 \pm 0.30) + (0.92 \pm 0.17) \times \log T_{peak}$ | 0.54 | $<10^{-4}$ |
| $T_{duration}$ ($T_{peak}$) | Afterglow Flares | $\log T_{duration} = (-0.28 \pm 0.11) + (1.06 \pm 0.05) \times \log T_{peak}$ | 0.71 | $<10^{-4}$ |
| $L_{p,iso}$ ($T_{peak,z}$) | Prompt Flares | $\log L_{p,iso} = (52.48 \pm 0.38) + (-1.29 \pm 0.21) \times \log T_{peak,z}$ | $-0.54$ | $<10^{-4}$ |
| $L_{p,iso}$ ($T_{peak,z}$) | Afterglow Flares | $\log L_{p,iso} = (52.50 \pm 0.22) + (-1.40 \pm 0.10) \times \log T_{peak,z}$ | $-0.64$ | $<10^{-4}$ |
| $L_{x,iso}$ ($T_{peak,z}$) | Prompt Flares | $\log L_{x,iso} = (52.38 \pm 0.40) + (-1.50 \pm 0.22) \times \log T_{peak,z}$ | $-0.59$ | $<10^{-4}$ |
| $L_{x,iso}$ ($T_{peak,z}$) | Afterglow Flares | $\log L_{x,iso} = (52.00 \pm 0.20) + (-1.38 \pm 0.09) \times \log T_{peak,z}$ | $-0.65$ | $<10^{-4}$ |
| $L_{x,iso}$ ($T_{duration,z}$) | Prompt Flares | $\log L_{x,iso} = (51.13 \pm 0.24) + (-0.83 \pm 0.14) \times \log T_{duration,z}$ | $-0.53$ | $<10^{-4}$ |
| $L_{x,iso}$ ($T_{duration,z}$) | Afterglow Flares | $\log L_{x,iso} = (51.18 \pm 0.16) + (-1.06 \pm 0.08) \times \log T_{duration,z}$ | $-0.69$ | $<10^{-4}$ |
| $L_{x,iso}$ ($L_{p,iso}$) | Prompt Flares | $\log L_{x,iso} = (-0.80 \pm 2.48) + (1.01 \pm 0.05) \times \log L_{p,iso}$ | 0.90 | $<10^{-4}$ |
| $L_{x,iso}$ ($L_{p,iso}$) | Afterglow Flares | $\log L_{x,iso} = (2.71 \pm 0.86) + (0.94 \pm 0.02) \times \log L_{p,iso}$ | 0.95 | $<10^{-4}$ |
| $E_{x,iso}$ ($S_F$) | Prompt Flares | $\log E_{x,iso} = (55.47 \pm 0.75) + (0.62 \pm 0.12) \times \log S_F$ | 0.57 | $<10^{-4}$ |
| $E_{x,iso}$ ($S_F$) | Afterglow Flares | $\log E_{x,iso} = (56.24 \pm 0.46) + (0.75 \pm 0.07) \times \log S_F$ | 0.67 | $<10^{-4}$ |
| $E_{x,iso}$ ($L_{x,iso}$) | Prompt Flares | $\log E_{x,iso} = (22.07 \pm 3.55) + (0.59 \pm 0.07) \times \log L_{x,iso}$ | 0.68 | $<10^{-4}$ |
| $E_{x,iso}$ ($L_{x,iso}$) | Afterglow Flares | $\log E_{x,iso} = (26.39 \pm 1.76) + (0.50 \pm 0.04) \times \log L_{x,iso}$ | 0.69 | $<10^{-4}$ |
| $E_{x,iso}$ ($L_{p,iso}$) | Prompt Flares | $\log E_{x,iso} = (15.82 \pm 3.28) + (0.71 \pm 0.07) \times \log L_{p,iso}$ | 0.79 | $<10^{-4}$ |
| $E_{x,iso}$ ($L_{p,iso}$) | Afterglow Flares | $\log E_{x,iso} = (25.17 \pm 1.53) + (0.52 \pm 0.03) \times \log L_{p,iso}$ | 0.79 | $<10^{-4}$ |
| $E_{x,iso}$ ($L_{p,iso}$) early | Prompt Flares | $\log E_{x,iso} = (15.87 \pm 3.25) + (0.71 \pm 0.06) \times \log L_{p,iso}$ | 0.79 | $<10^{-4}$ |
| $E_{x,iso}$ ($L_{p,iso}$) early | Afterglow Flares | $\log E_{x,iso} = (16.30 \pm 1.35) + (0.70 \pm 0.03) \times \log L_{p,iso}$ | 0.84 | $<10^{-4}$ |
| $S_F$ ($F_{peak}$) | Prompt Flares | $\log S_F = (0.54 \pm 0.76) + (0.86 \pm 0.09) \times \log F_{peak}$ | 0.76 | $<10^{-4}$ |
| $S_F$ ($F_{peak}$) | Afterglow Flares | $\log S_F = (-2.47 \pm 0.37) + (0.50 \pm 0.04) \times \log F_{peak}$ | 0.77 | $<10^{-4}$ |
| $S_F$ ($F_{peak}$) early | Prompt Flares | $\log S_F = (0.53 \pm 0.75) + (0.86 \pm 0.09) \times \log F_{peak}$ | 0.76 | $<10^{-4}$ |
| $S_F$ ($F_{peak}$) early | Afterglow Flares | $\log S_F = (-0.13 \pm 0.30) + (0.78 \pm 0.03) \times \log F_{peak}$ | 0.86 | $<10^{-4}$ |

**Note.** The correlations are derived from 69 prompt flares and 173 afterglow flares with redshifts. The subscript "early" indicates flares with peak times earlier than 1000 s. $R$ is the Spearman correlation coefficient and $P$ is the associated $p$-value.

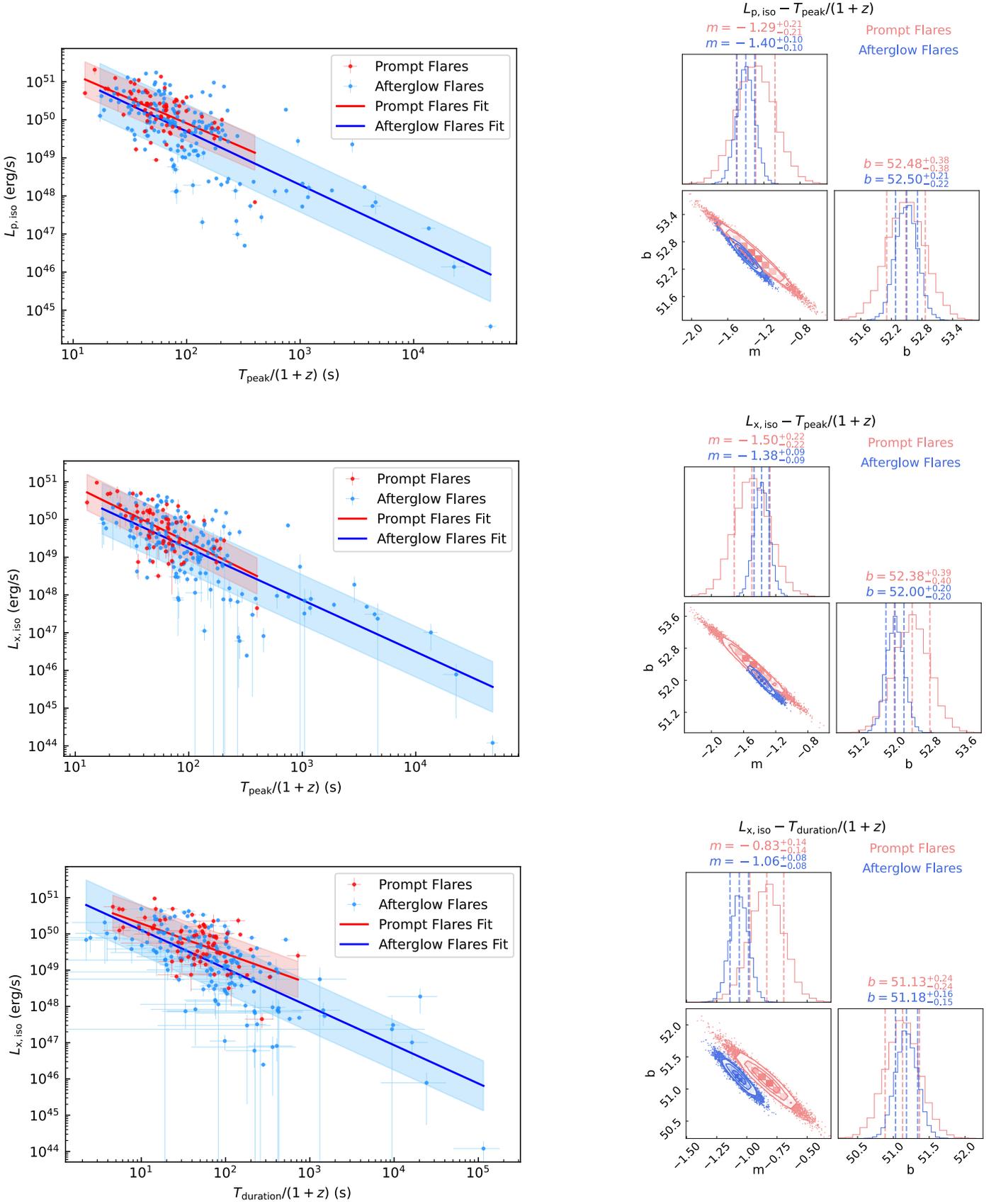

**Figure 9.** Top panel: the peak luminosity $L_{p,iso}$ is correlated with peak time $T_{peak}$. Middle panel: the average luminosity $L_{x,iso}$ is correlated with peak time $T_{peak}$. Bottom panel: the average luminosity $L_{x,iso}$ is correlated with the duration $T_{duration}$. $T_{peak}$ and $T_{duration}$ is converted to the source frame. Color conventions follow those in Figure 8.

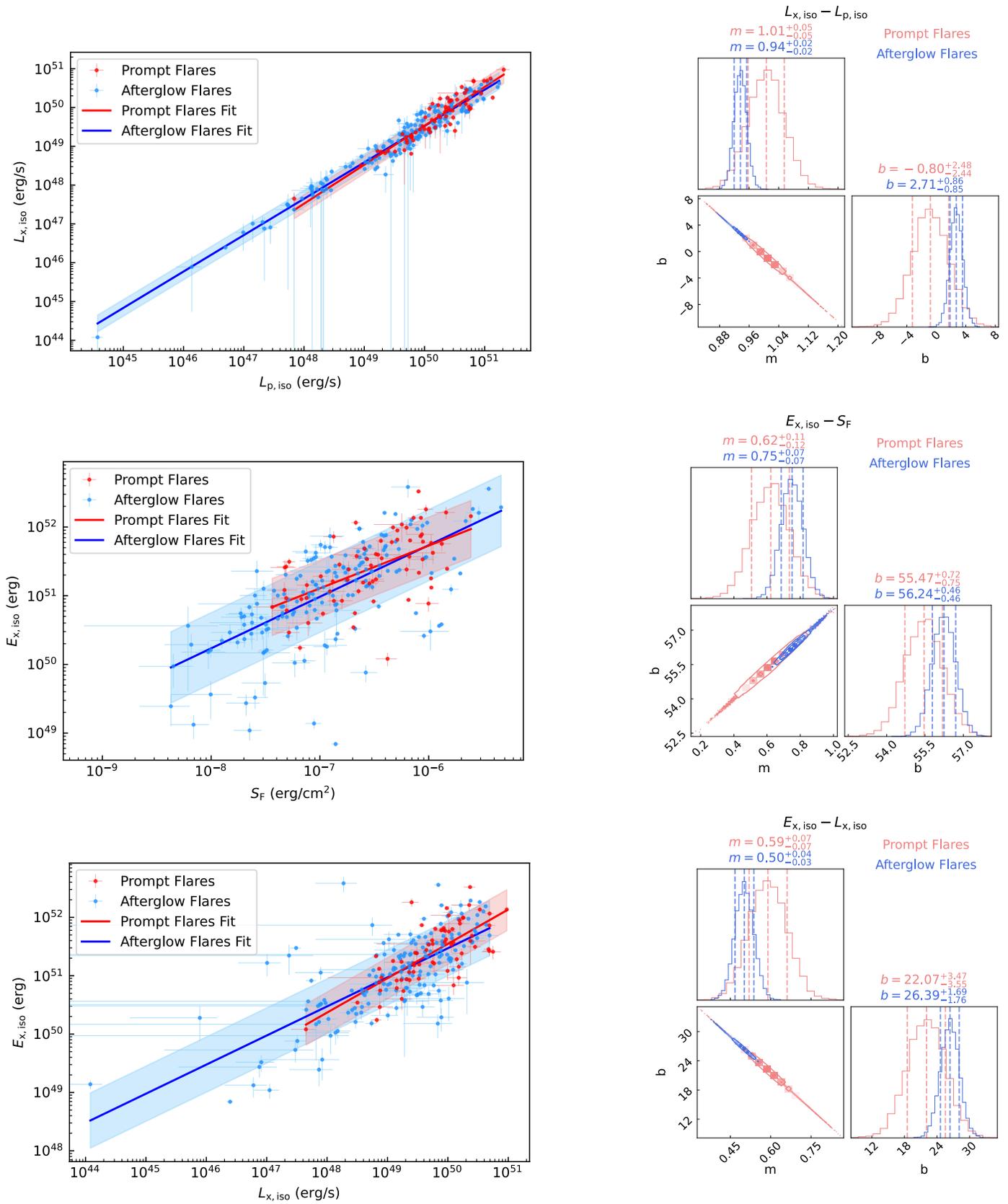

**Figure 10.** Top panel: the average luminosity $L_{x,iso}$ is correlated with peak luminosity $L_{p,iso}$. Middle panel: the isotropic energy $E_{x,iso}$ is correlated with fluence $S_F$. Bottom panel: the isotropic energy $E_{x,iso}$ is correlated with average luminosity $L_{x,iso}$. Color conventions follow those in Figure 8.

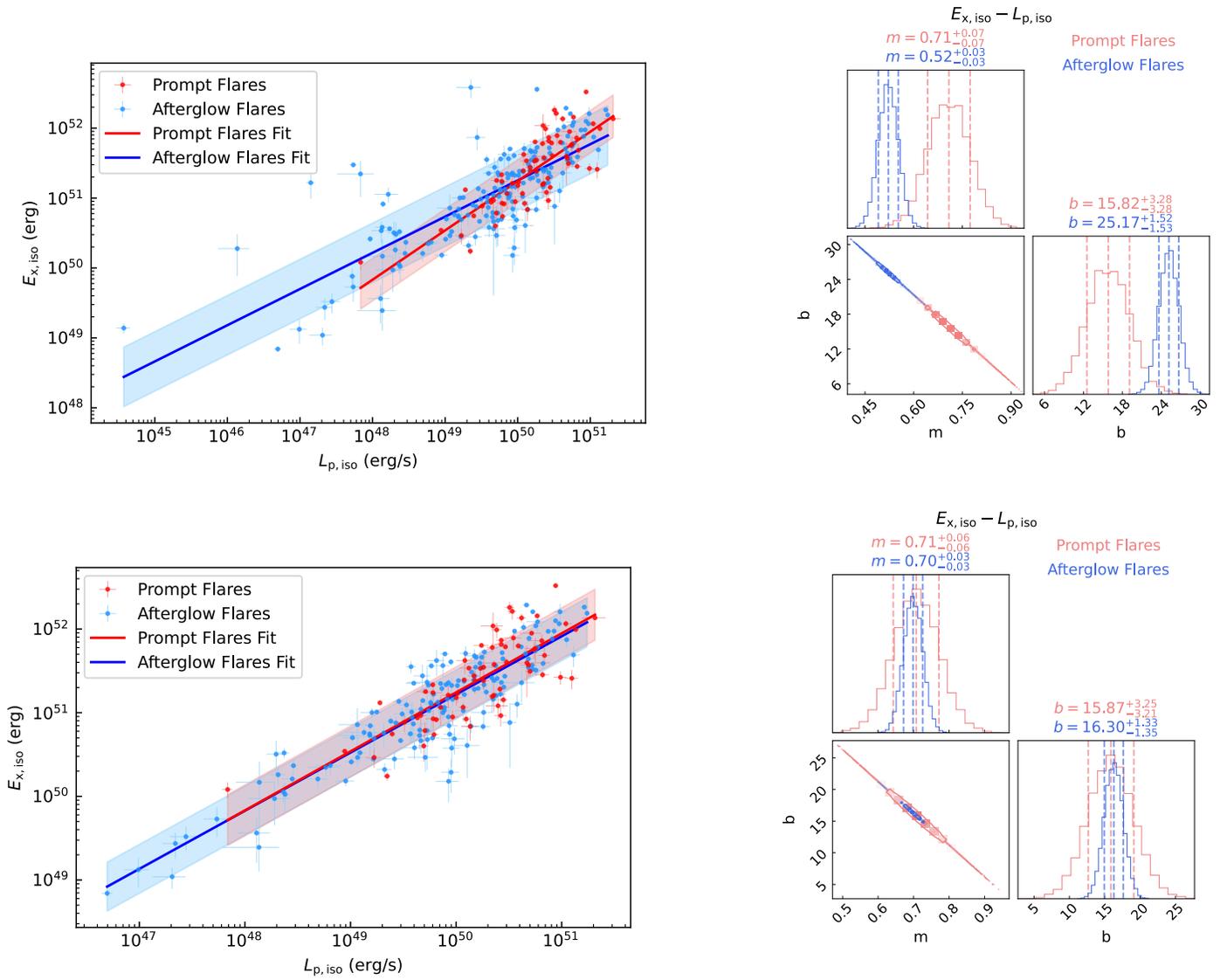

**Figure 11.** Top panel: for all flares with redshift, the isotropic energy $E_{\rm x,iso}$ is correlated with peak luminosity $L_{\rm p,iso}$. Bottom panel: for early flares with peak time earlier than 1000 s. Color conventions follow those in Figure 8.

## 4. Conclusions and Discussion

X-ray flares are frequently observed in GRBs, and they are generally believed to originate from the activity of the central engine, although direct evidence has been scarce. In this paper, we categorize X-ray flares into prompt flares ($T_{\rm peak} < T_{90}$ end time) and afterglow flares and attempt to determine the origin of afterglow flares by comparing their statistical similarities with prompt flares, which are more definitively linked to the activity of the central engine.

In this paper, we present a catalog of 701 luminous X-ray flares from 315 GRBs, sourced from the Swift/XRT data up to 2024 August. The flares are classified into 183 prompt flares (69 with measured redshifts) and 518 afterglow flares (173 with measured redshifts). Our statistical comparisons focus exclusively on the subsample of flares with confirmed redshifts, enabling a direct and physically meaningful analysis. The key findings for these redshift-corrected flares are summarized as follows:

(1) The morphological characteristics of the two types of flares are quite similar.

(2) For the entire sample of flares, the occurrence time of prompt flares precedes that of afterglow flares, and their luminosities and isotropic energy are higher than those of afterglow flares.

(3) For prompt flares and afterglow flares with identical temporal distributions, they not only share the same morphological characteristics but also exhibit similar luminosities and isotropic energy distributions.

(4) The correlations on the parameters of these two types of flares follow the same regularity.

Based on the above results, we can infer the following conclusions:

(1) The prompt X-ray flares and the afterglow X-ray flares originate from the same source, namely, the activity of the central engine.

(2) All X-ray flares follow the same statistical patterns, such as earlier X-ray flares being brighter and having shorter durations. This is consistent with the results of previous surveys (G. Chincarini et al. 2007, 2010; S.-X. Yi et al. 2016; Y.-R. Shi et al. 2022).

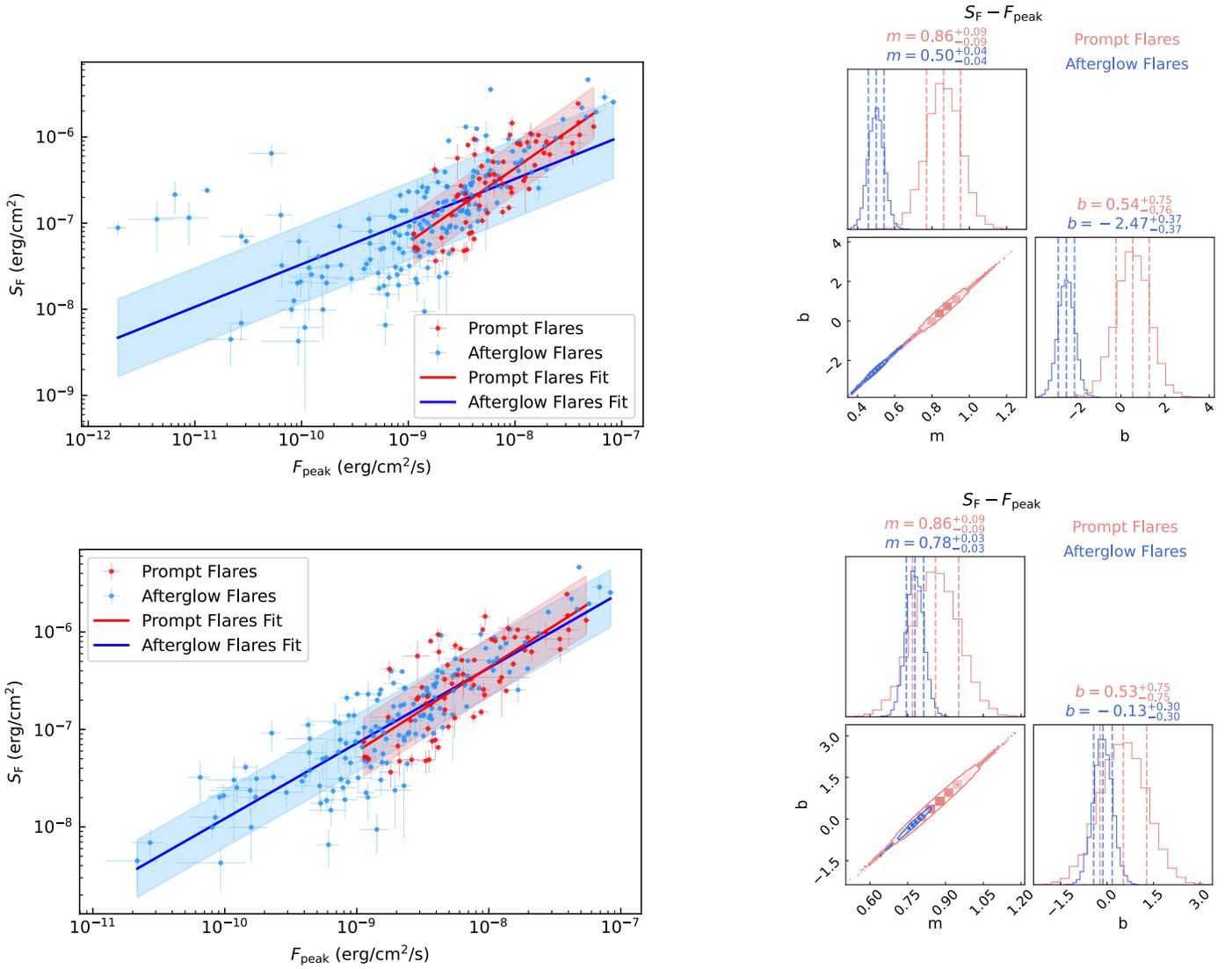

**Figure 12.** Top panel: for all flares with redshift, the fluence $S_F$ is correlated with the peak flux $F_{peak}$. Bottom panel: for early flares with peak time earlier than 1000 s. Color conventions follow those in Figure 8.

(3) The duration of the central engine activity in a GRB can be significantly longer than the timescale of the prompt gamma-ray emission. Therefore, when using the activity duration of a GRB to constrain the properties of the central engine and its progenitor, it is important to consider the characteristics of all X-ray flares and gamma-ray emission rather than focusing only on the prompt gamma-ray emission.

It should be noted that some early rapid decay afterglow background components (B. Zhang et al. 2006) cannot be distinguished from the decay phases of X-ray flares in our samples. Therefore, we did not fit these components as flares or underlying afterglow background. As a result, we may have missed some extremely early potential flares. However, we believe that with the operation of the Space-based multi-band astronomical Variable Objects Monitor (SVOM) and EP, the characteristics of extremely early X-ray flares and afterglow will be more comprehensively described. We have also noticed that among the bursts observed by SVOM and EP, a portion exhibits only X-ray emission without gamma-ray radiation. These X-ray-only events would be incorporated into our statistical sample to investigate potential differences in their distributions and characteristics compared to typical X-ray flares. Such analysis may provide valuable insights into the nature of GRB central engines and their progenitors.

## Acknowledgments

We thank the anonymous referee for the valuable comments and suggestions. This work made use of data supplied by the UK Swift Science Data Centre at the University of Leicester. This work is supported by the National Natural Science Foundation of China (Projects 12373040, 12021003) and the Fundamental Research Funds for the Central Universities.

## ORCID iDs

Yinuo Ma https://orcid.org/0009-0007-1390-7575
He Gao https://orcid.org/0000-0003-2516-6288

## References

Boër, M., Gendre, B., & Stratta, G. 2015, ApJ, 800, 16
Burrows, D. N., Romano, P., Falcone, A., et al. 2005, Sci, 309, 1833
Butler, N. R., & Kocevski, D. 2007, ApJ, 663, 407

Chang, X. Z., Peng, Z. Y., Chen, J. M., et al. 2021, ApJ, 922, 34
Chincarini, G., Mao, J., Margutti, R., et al. 2010, MNRAS, 406, 2113
Chincarini, G., Moretti, A., Romano, P., et al. 2007, ApJ, 671, 1903
Dai, Z. G., Wang, X. Y., Wu, X. F., & Zhang, B. 2006, Sci, 311, 1127
Evans, P. A., Beardmore, A. P., Page, K. L., et al. 2007, A&A, 469, 379
Evans, P. A., Beardmore, A. P., Page, K. L., et al. 2009, MNRAS, 397, 1177
Falcone, A. D., Burrows, D. N., Lazzati, D., et al. 2006, ApJ, 641, 1010
Falcone, A. D., Morris, D., Racusin, J., et al. 2007, ApJ, 671, 1921
Ford, L. A., Band, D. L., Matteson, J. L., et al. 1995, ApJ, 439, 307
Gao, H., Lei, W.-H., Zou, Y.-C., Wu, X.-F., & Zhang, B. 2013, NewAR, 57, 141
Gao, H., Ren, A.-B., Lei, W.-H., et al. 2017, ApJ, 845, 51
Gehrels, N., Chincarini, G., Giommi, P., et al. 2004, ApJ, 611, 1005
Gibson, S. L., Wynn, G. A., Gompertz, B. P., & O'Brien, P. T. 2017, MNRAS, 470, 4925
Gibson, S. L., Wynn, G. A., Gompertz, B. P., & O'Brien, P. T. 2018, MNRAS, 478, 4323
Hou, S.-J., Liu, T., Gu, W.-M., et al. 2014, ApJL, 781, L19
King, A., O'Brien, P. T., Goad, M. R., et al. 2005, ApJL, 630, L113
Kobayashi, S., Piran, T., & Sari, R. 1997, ApJ, 490, 92
Kouveliotou, C., Meegan, C. A., Fishman, G. J., et al. 1993, ApJL, 413, L101
Kumar, P., & Zhang, B. 2015, PhR, 561, 1
Lan, L., Gao, H., Li, A., et al. 2023, ApJL, 949, L4
Lazzati, D., Blackwell, C. H., Morsony, B. J., & Begelman, M. C. 2011, MNRAS, 411, L16
Lazzati, D., Perna, R., & Begelman, M. C. 2008, MNRAS, 388, L15
Lee, W. H., Ramirez-Ruiz, E., & López-Cámara, D. 2009, ApJL, 699, L93
Lei, W.-H., Zhang, B., & Liang, E.-W. 2013, ApJ, 765, 125
Liu, T., Liang, E. W., Gu, W. M., et al. 2010, A&A, 516, A16
Liu, Y., Sun, H., Xu, D., et al. 2025, NatAs, 9, 564
Lü, L.-Z., Liang, E.-W., & Cordier, B. 2022, ApJ, 941, 99
Luo, Y., Gu, W.-M., Liu, T., & Lu, J.-F. 2013, ApJ, 773, 142
Margutti, R., Guidorzi, C., Chincarini, G., et al. 2010, MNRAS, 406, 2149
Mészáros, P., & Rees, M. J. 1997, ApJ, 476, 232
Paczynski, B., & Rhoads, J. E. 1993, ApJL, 418, L5
Panaitescu, A., & Kumar, P. 2002, ApJ, 571, 779
Peng, F.-K., Hu, Y.-D., Wang, X.-G., Lu, R.-J., & Liang, E.-W. 2014, JApA, 35, 423
Perna, R., Armitage, P. J., & Zhang, B. 2006, ApJL, 636, L29
Romano, P., Moretti, A., Banat, P. L., et al. 2006, A&A, 450, 59
Saji, J., Iyyani, S., & Mazde, K. 2023, ApJS, 269, 2
Sari, R., & Piran, T. 1997, ApJ, 485, 270
Shi, Y.-R., Ding, X.-K., Zhu, S.-Y., Sun, W.-P., & Zhang, F.-W. 2022, Univ, 8, 358
Usov, V. V. 1992, Natur, 357, 472
Woosley, S. E. 1993, ApJ, 405, 273
Xie, W., Lei, W.-H., & Wang, D.-X. 2017, ApJ, 838, 143
Yi, S.-X., Xi, S.-Q., Yu, H., et al. 2016, ApJS, 224, 20
Yi, S.-X., Xie, W., Ma, S.-B., Lei, W.-H., & Du, M. 2021, MNRAS, 507, 1047
Yu, W.-Y., Lü, H.-J., Yang, X., Lan, L., & Yang, Z. 2024, ApJ, 962, 6
Zhang, B. 2018, The Physics of Gamma-Ray Bursts (Cambridge: Cambridge Univ. Press)
Zhang, B., Fan, Y. Z., Dyks, J., et al. 2006, ApJ, 642, 354
Zhang, B., & Mészáros, P. 2001, ApJL, 552, L35
Zhang, B.-B., Zhang, B., Murase, K., Connaughton, V., & Briggs, M. S. 2014, ApJ, 787, 66